\documentclass [11pt]{article}
\usepackage{setspace}
\usepackage{endnotes}
\usepackage{epsfig, amsfonts, amsmath, amssymb, amsthm, pictex, verbatim, amsthm, pdfsync, mathrsfs}
\usepackage{tensind}
\usepackage[square]{natbib}
\usepackage{endnotes}
\usepackage{fullpage}
\title{Einstein, the reality of space, and the action--reaction principle}
\author{ Harvey R. Brown\\
Faculty of Philosophy, University of Oxford\\ Radcliffe Humanities, Woodstock Road\\Oxford, OX2 6GG, U.K.\\{\em harvey.brown@philosophy.ox.ac.uk}\bigskip \\ 
Dennis Lehmkuhl \\ University of Wuppertal \\ IZWT and Institute of Philosophy \\ Gau{\ss}strasse 20, 42119 Wuppertal, Germany. \\ \emph{dennis.lehmkuhl@uni-wuppertal.de}}
\date{}

\begin{document}
	\newcommand{\conn}{\Gamma^{\rho}_{\hspace{1,5mm}\mu \nu}}
	\newcommand{\conna}{\Gamma^a_{\hspace{1,5mm}bc}}
	\newcommand{\riem}{R_{\mu \nu \sigma}^{\quad \omega}}
	\newcommand{\riema}{R_{abc}^{\quad d}}
	\newcommand{\ric}{R_{\mu \nu}}
	\newcommand{\weyla}{C_{abcd}}
	\newcommand{\weyl}{C_{\mu \nu \sigma \omega}}
	\newcommand{\M}{M_{\mu \nu}}
	\newcommand{\A}{A_{\mu}}
	\newcommand{\g}{g_{\mu \nu}}
	\newcommand{\gd}{\sqrt{-g}}
	\newcommand{\gk}{g_{AB}}
	\newcommand{\T}{T_{\mu \nu}}
	\newcommand{\R}{R_{\mu \nu}}
	\newcommand{\G}{G_{\mu \nu}}
	\newcommand{\F}{F_{\mu \nu}}
\bibliographystyle{agsm}

\maketitle

``\textit{For me it is an absurdity to ascribe physical properties to `space'.}" Albert Einstein to Ernst Mach,  December 1913.\footnote{Vol. 5, Doc. 495, The Collected Papers of Albert Einstein  (CPAE for short from now on). All translations from Einstein texts are based on the corresponding translation volumes of the CPAE unless otherwise noted; in some cases slight modifications of translation have been introduced.} 

\bigskip

``\textit{That a real thing has to be presupposed as the cause for the preference of inertial systems over non-inertial systems is a fact that physicists have only come to understand in recent years.}'' Einstein 1924.\footnote{\citet{einstein1991uber}, p. 88. Translation by DL.}

\begin{abstract}

Einstein regarded as one of the triumphs of his 1915 theory of gravity --- the general theory of relativity --- that it vindicated the action--reaction principle, while Newtonian mechanics as well as his 1905 special theory of relativity supposedly violated it. In this paper we examine why Einstein came to emphasise this position several years after the development of general relativity. Several key considerations are relevant to the story: the connection Einstein originally saw between Mach's analysis of inertia and both the equivalence principle and the principle of general covariance, the waning of Mach's influence owing to de Sitter's 1917 results, and Einstein's detailed correspondence with Moritz Schlick in 1920.
\end{abstract}

\section{Introduction}

It was striking in the 2012 IIAS seminar \textit{The Nature of Reality \ldots} that so many speakers independently referred to the quip by Albert Einstein in his 1930 dialogue with Rabindranath Tagore that he, Einstein, was more religious than the great Bengali polymath. Of course what Einstein was getting at is the fact that at the heart of the physical sciences, at least as he saw it, is a commitment to the reality of an external world whose goings-on are governed by laws that contain no fundamental reference to conscious agents, and in particular to human observers. For Tagore, all truth is human truth, if one is to take his claims literally.\footnote{For an insightful analysis of the Einstein-Tagore dialogues, see \citet{home1995einstein}.} For Einstein, ultimate truth about the physical world transcends the human realm. Einstein was a ``realist'', but his realism was of a modest, or non-metaphysical kind. It is the job of physicists, he argued, to come up with models of a mind-independent reality that are explanatory of our phenomenological experiences of natural regularities, within the laboratory and without. However, whether such models, when judged successful, correspond to the way the world ``really'' is, is a question Einstein thought best to leave aside. There are some philosophical questions for which Einstein thought the best response is a smile, and this was one of them. But Einstein stressed that even this weak, pragmatic take on truth involved a leap of faith. He made it clear, particularly in his 1949 \textit{Autobiographical Notes}, that Nature's connivance in allowing for the success in the scientific venture as he conceived it could not be a foregone conclusion.\footnote{See \citet{Einstein:1949}; for related remarks see \citet{einstein1948quanten}, reprinted and translated as ``Quantum mechanics and reality'' in \citet{einstein1971born}. Very helpful analyses of Einstein's philosophical position(s) can be found in \cite{howard1984realism}, \citet{howard1993einstein}, \citet{fine1986shaky}, \citet{vanDongen2010einstein}, chapter 2, and \citet{norton2012philosophy}.} It was conceivable for Einstein that mind, say, could be the bedrock of reality, but he felt that there was no good reason to start with that premiss, and good reasons not to. Realism for Einstein was more a program than a doctrine; it was a dogma about which he was careful not to be dogmatic.

Several years after the development of his 1915 general theory of relativity (GR), Einstein began to stress that physical space, or rather the metric field, not only constitutes a fundamental, autonomous element of objective reality, it plays a causal role in accounting for the inertial motion of bodies.\footnote{A discussion of the `principle of causality' is particularly prominent in the Einstein-Schlick correspondence as reviewed in section \ref{S:Schlick}. However, note that at the time it was quite common not to draw a clear distinction between causality and determinism; in many cases, demanding that the `principle of causality' holds is equivalent to demanding that every event has a determining cause or set of causes. For our discussion, the important point is that in the beginning 1920s Einstein started to think of the metric field as (causally) determining things, rather than just being determined by the distribution of masses.} He compared this with the active role of space  in the cases of Newtonian mechanics (NM) and special relativity (SR). In these cases, such putative action is clearly not reciprocated by bodies or fields: they do not act back on space-time structure, so the so-called action--reaction principle is violated. In contrast, in his relativistic theory of gravity GR, Einstein was to see the vindication of the principle. The metric can have a dynamical life of its own in the absence of matter fields (though, as we shall see, this goes against Einstein's original expectations) but, more to the point, when the latter exist, the metric affects and is affected by them.\footnote{\label{F:Lehmkuhl2011} In a Lagrangian framework (which Einstein started to use extensively from 1918 onwards), this mutual affection can be represented by the metric and the matter fields (both dynamical) \emph{coupling} to each other. The stress-energy tensor, however, turns out to be a relational property of the matter fields, which they posses in virtue of their relations to the metric field.  For a recent discussion of the relational significance of the stress-energy tensor, see \citet{Lehmkuhl2011}; section 4.3 for different kinds of coupling.}

Whether Einstein's reasoning is correct is open to doubt. What is debatable is not the claim that GR is consistent with the action--reaction principle, it is the claim that the older theories involving absolute space-time structures are not. More specifically, it is the assertion that such structures act in the relevant sense. Clearly, if they do not (as Newton himself argued regarding space; see below), then the fact that material bodies do not act back on them constitutes no violation of the action--reaction principle. The case for the view that absolute space-time structures of the kind that appear in NM and SR are not fundamental causal entities in their own right, but rather codifications of certain bare facts concerning the movement of bodies or behaviour of fields, has been made a number of times, in different ways.\footnote{For recent versions of this argument, see \citet{BrownPooley:2004} and \citet{Brown:2007}. It should be noted that  these analyses differ from Newton's, in the sense that there is no commitment to the reality of space as a fundamental entity.} One curious instance is arguably in a letter Einstein himself wrote to Ernst Mach in late 1913;\footnote{Vol. 5, Doc. 495, CPAE.} another is due to Moritz Schlick in correspondence with Einstein in 1920.\footnote{Schlick to Einstein, 10 June 1920, Vol. 10 Doc. 51 CPAE.} We shall return to the second case below; it turns out to be directly related to the main concern of this paper, which is why would Einstein start systematically emphasising the action--reaction principle in his defence of GR only in the 1920s.

\section{The action--reaction principle}

It is a venerable tradition in natural philosophy to assert that a substance is the seat of actions on other substances, and in turn subject to the actions of these other substances -- the action--reaction principle (AR). In his pre-\textit{Principia} manuscript \textit{De Gravitatione et Aequipondio Fluidorum}\footnote{A translation is found in \citet{hall1978unpublished}. The date of this important manuscript is still a matter of dispute. There is little doubt that it was written between 1666 and 1685; for details of the conflicting views on the likely date, and a defence of the early claim, see \citet{henry2011gravity}. } (or \textit{De Grav} for short), Newton insisted that natural philosophers tacitly understand substance as an entity that acts on things, even if they don't state this explicitly. He distinguished two kinds of action, one which stimulates the perceptions of thinking beings, and one between material bodies, as in collisions. (Later he would extend this second kind to action at a distance.\footnote{\textit{Ibid}.}) This distinction today seems of little import, and the fundamental premiss that it is interactions between systems that count in physics, and not their intrinsic properties, is close to the heart of the `structural realist' position that has been much discussed in recent years in the philosophy of science. Be that as it may, Newton is clearly appealing to a principle in the  \textit{De Grav} that is more fundamental and general than what he would later designate as his third law of motion in the \textit{Principia} -- though the latter is often referred to as the law of action--reaction. (We shall see shortly how space, for Newton, is a kind of exception to this fundamental principle.) 

Leibniz, whose views on the nature of space and time were so different to Newton's, nonetheless shared the same intuition. In fact, when defining substance as that which acts and can be acted upon, he understood he was adopting the view of the scholastics.\footnote{Leibniz made this point in a 1700 letter to Bernoulli; see \citet{gerhardt1856leibnizens}, volume III.} Caution must be exercised however in attributing AR, as it is standardly understood today, to Leibniz. In the light of his doctrine of pre-established harmony, the meaning of causation, or rather interaction, within his deep metaphysics is almost certainly at odds with those views adopted by the majority of current metaphysicians,\footnote{See for example \citet{ishiguro1977pre} and \citet{woolhouse1985pre}.} not to mention physicists --- and its scope is still controversial.\footnote{See for example \citet{brown1992there} for a careful discussion of the question whether pre-established harmony applies in Leibniz's philosophy at the level of aggregate substances.} 



If there is a questionable aspect of AR, it is less the claim that substances act (how otherwise could their existence be known to us?) than the notion that they are necessarily acted back upon, that action must be reciprocal. If all substances act, they do so in relation to other substances; these other substances therefore cannot be immune from external influences. Now it might seem arbitrary on \textit{a priori} grounds to imagine that the `sensitivity' of such substances is not universal. That is to say, it might seem arbitrary to suppose that not all substances react to others. But no such abstract qualms can be entirely compelling; Nature must have the last say. 

Nowadays, it is well-known in the foundations of physics that the de Broglie-Bohm (``pilot-wave'') interpretation of quantum mechanics, in its standard form, involves a dynamical agent (the wavefunction) that acts on corpuscles (point particles) but is not acted back upon, at least by the corpuscles. It is noteworthy that David Bohm himself clearly found such violation of AR uncomfortable in his (first) 1952 formulation of the theory, and attempted to remedy it, as have others after him.\footnote{See \citet{squires1994some}.} While defenders of the original pilot-wave theory can legitimately argue that AR is not a logical necessity, others see its violation as a blemish in the theory.\footnote{See \citet{brown1995reality}.} It seems fair to say that currently within the physics community there is no consensus supporting the failure of AR in quantum theory.

For his part, Einstein himself had already stated in 1922 that  it is ``contrary to the mode of scientific thinking to conceive of a thing . . . which acts itself, but which cannot be acted upon''.\footnote{\citet{einstein1922vier}.}
The object of Einstein's ire in 1922 was NM and his own creation, SR. Yet there is no hint in his writings around the time of the development of SR in 1905 that Einstein considered either of these theories to incorporate a violation of the action--reaction principle; at any rate the explicit condemnation came later. Why? In all probability because it was part of an honest sales pitch for GR, his greatest and most radical contribution to science, after Einstein was reluctantly forced to concede, because of results by de Sitter, that the theory as a whole was not consistent with ``Mach's Principle'', even though special solutions are. It seems that this change of tack on Einstein's part was consolidated in the mentioned 1920 correspondence with the physicist-philosopher Moritz Schlick.

But before examining the evolution of Einstein's views on the causal role of space-time, it is worth briefly visiting Newton's own views on the nature of absolute space, especially as expounded in the  \textit{De Grav}. This manuscript contains a hard-hitting critique of Descartes' relational theory of motion, and the positive reasons why Newton felt compelled to posit the existence of absolute space; these important lines of reasoning have been discussed extensively in the literature.\footnote{See \citet{Stein:1967}, \citet{barbour1989absolute}, Chapter 11, \citet{rynasiewicz1995Scholium1, rynasiewicz1995Scholium2}, and \citet{Pooley:forth}, Chapter 2.} What concerns us is Newton's insistence that space, despite its reality, does not act on either our sense organs (which is patent) or on bodies. If it is a substance, it is by Newton's own reasoning \textit{sui generis} in its causal inefficacy, and ultimately Newton had no interest in classifying it as either substance or accident.\footnote{\textit{de Grav} (see note 7 above), section 7. The inspiration for Newton's view that space is neither substance nor accident may have been the Renaissance thinker Francesco Patrizi da Cherso, or indeed Walter Charleton. See in particular \citet{Pooley:forth}, section 2.1, for a discussion of possible influences on Newton in this regard.  A translation of and commentary on Patrizi's ``De Spacio Physico'' can be found in \citet{brickman1943physical}.} 

When, in the context of discussions relevant to this essay, Newtonian space is assigned a causal role, it is usually to account for inertia, \textit{i.e.} the privileged existence of inertial frames, or equivalently the special motions of force-free bodies. In the \textit{De Grav}, Newton explicitly renounced such a role. He stated that the reason why projectiles that are not being acted upon by other bodies move in straight lines and uniform speed is precisely that space has no ability to help or hinder any change in the motion of bodies!\footnote{\textit{de Grav} (see note 7 above), section 5.}

\section{Einstein on absolute space}

\subsection{Einstein and Mach}

Einstein's tortuous road to the 1915 field equations of his unhappily-named general theory of relativity followed a number of fundamental, partially overlapping, conceptual signposts, apart from the requirement of securing the Newtonian limit in the case of suitably defined weak gravitational fields. These were the so-called equivalence principle, which connects gravitational effects with inertia; Mach's ideas concerning the origins of inertia; the principle of the relativity of motion, and finally the conservation of energy-momentum. It is probably no exaggeration to say that none of these guiding principles was to survive fully intact once the promised land was reached.  In this section we shall be concerned in particular with Mach's influence on Einstein's thinking,  because its demise is intimately connected with the appearance of AR in Einstein's arsenal of arguments in favour of GR. The story of Einstein's debt to Mach has of course received considerable attention in the literature;\footnote{See particularly \citet{Barbour:1990}, \citet{norton1995mach}, \citet{Hoefer:1995,Hoefer:1994}, \citet{Renn:2007}; and see \citet{Lehmkuhl:forth_Mach}. \label{F:Mach Sources}} we shall deal only with those elements that are necessary for our ends.


\subsection{Mach}

Ernst Mach's own thinking on inertia is usefully divided into strands: his critique of Newtonian mechanics, more specifically of absolute space and time, and his somewhat vague and varied suggestions as to how to remedy the theory. As for the first strand, Mach's central objection to NM is that in appealing to motion with respect to absolute space in accounting, say, for centrifugal effects in rotating bodies, one is being unfaithful to the fundamental aim of science, which has to do with providing an economic systematization of experience. Space is not observable, so a purportedly fundamental epistemological (or perhaps better methodological) principle requires us, according to Mach, to refer only to motion relative to other bodies. It is not entirely clear that Mach sees space as `acting' on a rotating body within the standard Newtonian account; and anyway there appears to be no hint of a complaint in Mach's writings that it violates the action--reaction principle. The issue is the methodological/epistemological role of observables in physics associated with Mach's brand of empiricism, not the metaphysics of action.

Further insight into Mach's thinking is seen in the second, positive strand. Opinion is divided as to whether he was suggesting a reformulation of Newtonian mechanics or a new, distinct theory accounting for the inertial motion of bodies. John Norton, for example, is skeptical about the seriousness of Mach's revolutionary intentions;\footnote{\citet{norton1995mach}.} Julian Barbour is not. Barbour indeed provides powerful textual evidence that Mach, on at least one occasion, was searching for an explanation of the inertia (as opposed to inertial mass) of a force-free body, i.e. its uniform, rectilinear motion, in terms of distant bodies, in analogy with the explanation of the acceleration of a body resulting from the gravitational influence of distant masses -- an account which would yield the Newtonian predictions to a good approximation in the case of a universe populated to the extent ours is.\footnote{\citet{barbour1995general}, section 2. Note that this paper also defends a version of Machianism that, unlike Mach's own version, is consistent with (certain solutions of) the field equations of GR.} 

There are three points to be made at this stage. First, if Mach's concern were (counterfactually) truly AR, would it be beyond the bounds of reasonableness to suppose he might have sought to make space dynamical, and capable of being acted on by bodies? There is no hint of this possibility in Mach's writings, despite the fact that dynamical notions of space had already been proposed in 1872 by Z\"{o}llner and independently in 1876 by Clifford.\footnote{Both of these figures were interested in non-Euclidean geometry, and eschewed Riemann's separation of physics from geometry. (For further discussion of their views, see \citet{kragh2012geometry}, \citet{kragh2012zollner}, where it is pointed out that in 1872 Z\"{o}llner proposed a cosmological model describing a finite universe in closed space. A connection between Z\"{o}llner's suggestion of a cosmological underpinning of the force law in electrodynamics and ``later versions of Mach's principle'' is suggested in \citet{kragh2012geometry}, p. 26.) It is unclear to us, however, just to what extent these views are fully consistent with the action--reaction principle.} Mach in fact showed considerable interest in non-Euclidean geometry, and thought it probable, but not inevitable, that it would have no role to play in physics. If it did, it would for Mach only be through the behaviour of observable, material bodies, and there was no evidence for it when he wrote essays for \textit{The Monist} which would be published collectively as \textit{Space and Geometry} in 1906. He thought that the behaviour of matter was as unlikely to indicate the reality of non-Euclidean geometry as it would satisfy ``the atomistic fantasies of the physicist.''\footnote{Mach (1906), p. 136 and p. 141. See in this connection \citet{kragh2012geometry}, p. 35.}

Second, it would seem that Mach's gravitation-like proposal for the origin of inertia, and the very existence of inertial frames, would involve action at a distance.\footnote{Indeed, in 1920, Einstein would explicitly reject Mach's original reasoning as incorporating action at a distance; see \citet{Einstein:1920g}, p. 11. He would repeat the objection in \citet{einstein1991uber}.}  In fact, it could be called super-action-at-a-distance; the inertial motion of a body is being attributed to the existence of celestial bodies \textit{so far} away that their gravitational actions on it are negligible. The caveat is that Mach's own notion of causality was rather thin; Norton has called it ``idiosyncratic''.\footnote{\citet{norton1995mach} p. 28.} Mach saw physics as providing only functional dependencies between experiences; systematic correlations rather than causal interactions (in so far as the distinction is meaningful). The commitment to the notion that ``\textit{the law of causality is sufficiently characterised by saying that it is the supposition of the mutual dependence of phenomena}'' on Mach's part\footnote{\citet{mach1872geschichte}, p. 61; Mach's emphasis.} perhaps explains why he was comfortable enough with Newton's picture of gravity to encourage the search of an analogous account of inertia. However all this may be, it seems reasonable to conclude that Mach was not obviously concerned, in the context of his analysis of inertia in NM, with the action--reaction principle or its violation, in anything like the ordinary sense. (It is true that Mach provided an operational reading of inertial mass based on Newton's third law of motion, one that proved to be influential. But as we have urged, the third law is not to be conflated with the AR principle.) In fact, a degree of resonance is discernible between Mach's view of the nature of causal connections and that of Leibniz, despite the chasm between their views on metaphysics.

The third point is closely related to the second. An important aspect of Mach's thinking about inertia is the emphasis on the cosmological nature of its origins. It is not some subset of distant bodies that determines the system of inertial frames; it is the totality of bodies in the universe. Barbour has aptly connected this cosmological strand of Mach's reasoning -- the requirement of \textit{self-referentiality} in any adequate account of the observed world -- with Kepler's 1609 theory of place and motion.\footnote{\citet{Barbour:1990}, p. 48.} Prior to 1917, the strand played only a minor role in Einstein's thinking.

\subsection{Back to Einstein}

Einstein's decade-long love affair with Mach's philosophy of inertia was complicated and tortuous. The first complication is that Einstein entertained  two quite distinct Mach-inspired doctrines, one of which actually had no basis in Mach's writings, as Barbour first emphasised in 1990.\footnote{\textit{Op. cit.}, pp. 49-50. For further analysis, see in particular \citet{Hoefer:1995} and \citet{Lehmkuhl:forth_Mach}.} This was the doctrine that the \textit{inertial mass} of a body is to be explained as arising from the presence of other bodies, with the consequence that a body at spatial infinity should have zero mass. (Mach himself had no difficulty in viewing inertial mass as an intrinsic property of the body, and, as mentioned above, used Newton's third law to reveal its operational significance.)  Indeed it was this idea of the ``relativity of inertia'' that Einstein had in mind in his first endorsement of Mach's reasoning in a paper published in 1912.\footnote{\citet{Einstein1912e}.}


It has to be noted that this paper was written at a remarkable moment of time. In March 1912, Einstein had completed his final paper on a theory of static gravitational fields in which gravity was represented by a scalar $c$, which was supposed to represent not only the gravitational potential at the space-time point in question but also the speed of light at that very space-time point.\footnote{Einstein finally abandoned  this theory primarily because he convinced himself that the theory was in conflict with Newton's third law, see \citet{Einstein:1912d}, p. 452-458; and see \citet{norton1995eliminative}, section 5.1, for details. \label{F:Newton3}} In August 1912, Einstein moved from Prague to Zurich, where his collaboration with Marcel Grossmann began, one which culminated in  the first paper\footnote{\citet{EinsteinGrossmann:1913}.}, propounding what is known as the \textit{Entwurf} theory, in which gravity is represented by a dynamical ten-component metric tensor field $\g$ whose non-dynamical counterpart Einstein already knew from  Minkowski's 1908 formulation of special relativity. Thus,  the cited 1912 paper was written only a few months before Einstein learned about the powers of the metric tensor, and it is noteworthy that he had already endorsed the idea of a relativity of inertia in the context of a relativistic theory of gravity.

Einstein states in the 1912 paper\footnote{\citet{Einstein1912e}, p. 38.} that the results he has obtained give support to the (alleged) idea of Mach that the inertia of point masses is a result of the presence of other masses, that it rested on an \emph{interaction} (\emph{Wechselwirkung}) of the point particle with those other masses.  (Recall that Mach himself did not use the term `interaction' but spoke of mutual dependencies.\footnote{See \citet{wolters1987mach} for a comparison of Mach's and Einstein's choice of words, respectively.}) We find similar statements in the mentioned 1913 paper with Grossmann and another 1913 paper by Einstein\footnote{\citet{Einstein:1913c}, p.1260.}; in each instance Einstein emphasises that the inertia of a body should be derived as the result of an interaction of this body with other bodies.\footnote{Note that Einstein did not clearly distinguish between the relativity of inertia (the predecessor of `Mach's principle' as defined only in \citet{Einstein:1918}, and the relativity of motion, as he himself admits of not having done up until \citet{Einstein:1918}). For details on different versions of these principles and the development in Einstein's thought see the sources summarised in footnote \ref{F:Mach Sources}.} Also in 1913, 
Einstein wrote two letters to Mach, the first in June, the second in December.\footnote{See Vol.5, Docs. 448 and 495, CPAE.} In the June letter, Einstein writes enthusiastically that in his \textit{Entwurf} theory which he developed with Marcel Grossmann, ``\textit{inertia} has its origin in some kind of \textit{interaction} of the bodies, completely in the sense of your considerations about Newton's bucket experiment.'' Despite the wording, careful reading of the letter makes it clear that again Einstein is thinking of inertial mass rather than the inertial frame.\footnote{In the letter, this becomes clear when Einstein directs Mach's attention to a particular page in the \emph{Entwurf} paper (p. 6), where he explicitly claims that the inertial mass of a body is a function of the gravitational potential in the static limit of the \emph{Entwurf} theory, which he then relates to Mach's ``bold idea'' of a relativity of inertia.}

However, in a 1914 paper\footnote{\citet{Einstein:1914o}, p. 74.}, Einstein explicitly discussed Newton's famous bucket thought experiment and Mach's criticism thereof --- arguing that if we have a choice, we should go for an account in which no absolute motion exists. Here, though not for the first time, Einstein is advocating something much closer to Mach's own concerns, and it was to dominate his thinking for at least the next two years, as we shall see. At this point, it is worth noting that in his well-known 1999 study of the principle of general covariance, John Norton argued that as early as 1913, Einstein articulated a connection between Mach's theory of inertia as he understood it and the action--reaction principle, AR. Indeed, Norton sees concern with the violation of AR in pre-GR space-time theories as ``the enduring core of the cluster of ideas that led Einstein to the relativity of inertia and Mach's principle''.\footnote{\citet{norton1999general}, p. 810.} We shall defer discussion of this claim to section \ref{S:Whence AR}.

\subsection{Hope of a solution?}

How was the Machian positive program related primarily to inertial motion (rather than to inertial mass) to be implemented? Essentially by finding a way to bypass it! Einstein was to link the problem of inertial motion with a notion he expressed clearly in 1911, itself related to the equivalence principle: that a uniformly accelerating reference frame (which reproduces all the effects of a homogenous gravitational field) is no more absolute than an inertial frame. Indeed, he hoped that his future theory of gravity would allow for a yet further generalisation of this putative extension of the Galilean-Einstein relativity principle -- to \textit{all} frames, such that the very distinction between inertial and non-inertial motion would become relative, non-absolute.\footnote{See \citet{norton1999general}, section 3.1.} By 1912, Einstein was convinced that the success of the complete ``relativity of motion'' would be guaranteed if the gravitational field equations turned out to be generally covariant.\footnote{When Einstein failed to find convincing generally covariant field equations, he convinced himself of their impossibility with the now famous `hole argument', only to return to the requirement of general covariance in launching the `point-coincidence argument' in 1915. See \citet{Stachel:1989}, \citet{EarmanNorton:1987} \citet{norton1999general} and \citet{Pooley:forth} for details, including references to criticism of Einstein's equating the `general principle of relativity' with general covariance.} What is important for our purposes is that Einstein saw relativity of inertia, the principle of the relativity of motion and the equivalence principle as walking hand in hand. As Barbour has stressed:
\begin{quote}
The drift of Einstein's thought is now clear. Whereas the logic of Mach's comments called for explicit derivation of the distinguished local frames of reference from a relational law of the cosmos as a whole, Einstein is working towards elimination of the problem of the distinguished frames by asserting that they are not really distinguished at all.\footnote{\citet{Barbour:1990}, p. 53.}
\end{quote}

Renn and Sauer have also emphasised essentially the same point (though with less emphasis on the departure from Mach's cosmological speculations):
\begin{quote}
Einstein's view that it made sense to search for a generalization of the relativity principle of classical mechanics and special relativity was \ldots based on his acceptance of a philosophical critique of classical mechanics raised by Mach and others. According to this critique, the justification of the privileged role of inertial frames of reference by the notion of absolute space was problematic, while the inertial forces experienced in accelerated frames of reference require an explanation in terms of the interaction between physical masses. Such an explanation would then eliminate any need for absolute space as a causal agent in the analysis of motion. The generalized relativity principle would go, so at least was Einstein's expectation, a long way, and might actually go all the way, towards an implementation of Mach's critique of classical mechanics in the new theory of gravitation.\footnote{Renn and Sauer, p. 301.}
\end{quote}
The culmination of this reasoning would appear in Einstein's great review paper of 1916.

\subsection{1916}
At the start of his 1916 review paper on the new general theory of relativity,\footnote{\citet{Einstein:1916e}.} Einstein would provide several (too many?) motivations for the requirement of general covariance; one of them was Mach's analysis of inertia. He went on to introduce a famous thought experiment involving two fluid (i.e. not entirely rigid) bodies $S_1$ and $S_2$ in empty space rotating around each other at constant angular velocity, and so far apart and isolated that the gravitational force between them and all other bodies is negligible.  Suppose it is a verifiable fact that body $S_1$  is spherical in shape, while body $S_2$ is ellipsoidal. Einstein asserts that, in answering the question as to why this is so, 
\begin{quotation}
No answer can be admitted as epistemologically satisfactory, unless the reason given is an \textit{observable fact of experience}. The law of causality has not the significance of a statement as to the world of experience, except when \textit{observable facts} ultimately appear as causes and effects.

Newtonian mechanics does not give a satisfactory answer to this question. It pronounces as follows: The laws of mechanics apply to a space $R_1$, in respect of which the body $S_1$ is at rest, but not to a space $R_2$, in respect to which the body $S_2$ is at rest. But the privileged space $R_1$ of Galileo, thus introduced, is merely a \textit{factitious} cause [\emph{fingierte Ursache}], and not a thing that can be observed. It is therefore clear that Newton's mechanics does not really satisfy the requirement of causality in the case under consideration, but only apparently does so, since it makes the factitious cause $R_1$  responsible for the observable [shape] difference in the bodies $S_1$  and $S_2$.
\end{quotation}

It is clear in this passage that despite Einstein's talk about the ``law of causality'' -- for which Mach's understanding would probably be very different from Einstein's -- his critique of NM is very similar to Mach's, and it is explicitly described as epistemological in nature. Indeed, the ``epistemological'' shortcoming of NM had already been mentioned by Einstein in 1913.\footnote{See \citet{Hoefer:1995}, p. 294.} But a notable feature of Einstein's Machianism is its selectivity. Mach rejected the existence of atoms on essentially the same grounds that he rejected the existence of absolute space; Einstein did not. By 1905, Einstein had made important contributions to the theory of capillarity, molecular dimensions, and statistical mechanics --- not to mention the nascent quantum theory. His 1905 paper on  Brownian motion  in particular would provide the basis for arguably the first significant empirical evidence of the existence of atoms. As Don Howard has written:
\begin{quote}
A careful reading, especially of Einstein's papers on the foundations of statistical physics, reveals that the influence of Mach and Ostwald was being felt. It was not, however, in the form of any doubt about the reality of atoms, but in the form of a caution about prematurely investing these atoms with any properties other than those necessary for the purpose at hand. \ldots It was thus not an ontology of unobservables that troubled Einstein; it was merely an ontology that was richer than it need be.\footnote{\citet{howard1993einstein}, p. 211.}
\end{quote} 

But the case of immaterial space was different. In his 1916 paper Einstein went on to claim that the proper cause of the difference between the bodies $S_1$  and $S_2$ must be sought in the relations between $S_1$, $S_2$ and other physical bodies, rather than in their relations to unobservable absolute space.
\begin{quotation}
These distant masses (and their motions relative to $S_1$ and $S_2$) must then be regarded as the seat of the causes (which must be susceptible to observation) of the different behaviour of our two bodies $S_1$ and $S_2$. They take over the role of the factitious cause $R_1$. Of all imaginable spaces $R_1$, $R_2$, etc., in any kind of of motion relatively to one another, there is none which we may look upon as privileged \textit{a priori} without reviving the above-mentioned epistemological objection. \textit{The laws of physics must be of such a nature that they apply to systems of reference in any kind of motion.} Along this road we arrive at an extension of the relativity principle.\footnote{\citet{Einstein:1916e}.}
\end{quotation}

The connection in Einstein's thinking between the problem of inertia and the generalised relativity principle is clearly stated here. In the previous year he had regained confidence in the general covariance of the gravitational field equations (following the notorious struggle with  his ``hole argument'') and thus found a purported solution to the Machian problem simply on the basis of the fact that the 1915 field equations are generally covariant. It is ironic then that for most of the 1916 paper, Einstein reverts to the use of unimodular coordinates (for which $\sqrt{-g} = 1$, $g$ being the determinant of the metric field $g_{\mu\nu}$) with the explicit purpose of simplifying the equations and calculations.\footnote{See \citet{Brown:2007}, Appendix A. Einstein's views on the meaning of general covariance would undergo considerable changes. Already in \citet{Einstein:1916b}, he would emphasise its role in clarifying  the conservation principle in GR, in providing what was a special case of Noether's second 1918 theorem; see \textit{ibid} and especially \citet{Rennjanssen:2006}, section 3. See also \citet{norton1999general}, especially section 5, for the development of the role of general covariance in Einstein's thought.} Be that as it may, it is worth noting, as Barbour did in 1990,\footnote{\citet{Barbour:1990}.} that in so far as general covariance was to play a role in solving the Machian conundrum, it was the relativity of motion, and not the relativity of inertial mass, that was relevant. 

1916 also marked the year of Mach's death. Einstein wrote an obituary, in which he discussed in detail Mach's criticism of Newton's concept of absolute motion, which in turn rested on the notion of absolute space. Einstein wholeheartedly subscribed to Mach's point that physics should avoid using these absolute concepts, and even makes fun of the ``philosophers'' with their ``treasure chests of the `absolute' and the `a priori'''.\footnote{\citet{einstein1916orbituarymach}.}


\subsection{Mach in trouble: 1917 to 1921}

Yet Einstein in 1916 had not yet relinquished his 1912 interpretation of Mach's philosophy in terms of inertial mass.  Einstein and Grossmann had already derived a related effect in the \textit{Entwurf} theory\footnote{See \citet{EinsteinGrossmann:1913}, p. 6}, namely that a mass point's inertial mass grows if other bodies with inertial mass are close by. In his 1921 Princeton lectures,\footnote{\citet{einstein1922vier}, p.64-66.} after de Sitter's attack on Mach's principle as defined in 1918 (see below), Einstein considered the limit where the geodesic equation goes over to the Newtonian equation of a mass point subject to a gravitational field, where $g_{\mu \nu} = \eta_{\mu \nu} + \gamma_{\mu \nu}$ and $\gamma_{\mu \nu}$ corresponds to  the Newtonian gravitational potential assumed to be small to first order.  He derived a form for the equations of motion which he interpreted as showing that, in GR too, the inertial mass of a mass point increases if other massive bodies are close by. This result he took  as ``strong support for Mach's ideas as to the relativity of all inertial actions''. He went even further:
\begin{quote}
If we think these ideas consistently through to the end we must expect the \emph{whole} inertia, that is, the \emph{whole} $\g$-field, to be determined by the matter of the universe [...] .  

\end{quote}	

Einstein then showed that this can indeed be achieved for the (static, spatially closed) Einstein universe. 

Of course, this was already a fall-back position. Originally, Einstein had expected that in \emph{every} solution of the Einstein equations, i.e. in every possible universe, the mass-energy of matter $\T$ would uniquely determine the $\g$-field. De Sitter brought an end to this idea, as will be described below.

Either way, taking the (partial) relativity of inertial mass of a body as reason to believe that the $\g$-field should be entirely determined by the $\T$-field is a remarkable conceptual jump. But this notion of $\g$ supervening on $\T$ was far from new in Einstein's Mach-inspired thinking; it can be traced back to the \textit{Entwurf} theory. Later, in 1918, it was taken by Einstein to \textit{define}  ``Mach's Principle". What then had happened to his notion of linking Mach's philosophy with general covariance?

The issue had become much more complicated. Earlier in 1918, Kretschmann would famously question whether general covariance had any empirical content at all.\footnote{See \citet{kretschmann1918}.} Kretschmann's analysis forced Einstein to clearly distinguish the relativity of motion principle from both the equivalence principle and from the view that the gravito-inertial field represented by the metric tensor must be determined exclusively by the distribution of masses in the universe. In 1918, his precise formulation of ``Mach's Principle'' became the claim that the $g_{\mu\nu}$ field must be ``conditioned and determined'' by the mass-energy-momentum $T_{\mu\nu}$ of matter.\footnote{\citet{Einstein:1918}.}

Alas, although no one seems to have appreciated  it at the time, $T_{\mu\nu}$ cannot be defined independently of $g_{\mu\nu}$.\footnote{This was noted in footnote \ref{F:Lehmkuhl2011} above; the dire implications for the 1918 Mach's Principle were spelt out recently in \citet{Lehmkuhl2011}, p. 482.} Einstein himself recognised this point explicitly in 1954, the year before his death; in a letter to Felix Pirani he admitted that Mach's Principle was hopeless for exactly this reason.\footnote{See \citet{BarbourPfister:1995}, p. 93.}  However, even in 1918, other grounds for doubting the validity of the principle already existed. A solution of Einstein's 1915 field equations is a static metric field corresponding to flat, matter-free Minkowski space-time; this was one reason why in 1917 Einstein felt he had to introduce modified field equations with a cosmological constant to rescue the principle.\footnote{\citet{Einstein:1917b}.} Alas again, de Sitter would soon show that the modified field equations admit other vacuum solutions: the degrees of freedom of the metric field are in general not uniquely determined by the mass-energy of matter.\footnote{Einstein initially tried to attack de Sitter's solution on various grounds, arguing that it was i) not static (unacceptable to Einstein at the time) and ii) that it involved an intrinsic singularity and thus matter. In the end, Einstein hat to admit that it was an entirely viable mathematical solution to the modified field equations (involving no intrinsic singularities). He went on to rule  out the solution as unphysical because it was not globally static; a move he repeated when \citet{friedman1922krummung} published his non-static solution and regretted painfully after \citet{hubble1929relation} published his results on the redshift of galaxies. See \citet{frenkel2002einstein} for Einstein's attack on Friedman's work and subsequent developments,  \cite{einstein1932relation} for Einstein's reaction to Hubble's results, and the Editorial note `The Einstein-de Sitter-Weyl-Klein debate' of Volume 7 CPAE for more on the exchange between Einstein and de Sitter, as well as \citet{janssen2008no}.} In the years between 1918 and 1922, Einstein was thus forced to admit the metric field $\g$ as a dynamical player in its own right according to GR, akin to the electromagnetic field $\F$. Nonetheless, he still clung to the weak form of Mach's principle present in the Princeton lectures, which holds only in the Einstein universe -- supposedly the best model of the actual universe.

These lectures are also the place where Einstein for the first time published an explicit, clear connection between the Machian critique of absolute space and the action--reaction principle AR. Here, he  stated:

\begin{quote}
[I]t is contrary to the mode of thinking in science to conceive of a thing (the space-time continuum) which acts itself but which cannot be acted upon. This is the reason why E. Mach was led to make the attempt to eliminate space as an active cause in the system of mechanics. [...] In order to develop this idea within the limits of the modern theory of action through a medium, the properties of the space-time continuum which determine inertia must be regarded as field properties of space, analogous to the electromagnetic field.\footnote{\citet{einstein1922vier}, p. 36.} 
\end{quote}  

This is hardly a faithful account of Mach's reasoning. More to the point, we witness here what seems to be an important shift in Einstein's  thinking about the nature of the problematic role space(time) plays in Newtonian mechanics and special relativity. But the seeds for this shift were planted at least two years earlier, as we shall now see.

\section{The Einstein-Schlick correspondence}
\label{S:Schlick}

\subsection{The background}

The growing recognition, on Einstein's part, of the tension between the field equations in GR and his 1918 version of Mach's Principle led him, as we have seen, to effectively assign genuine degrees of freedom to the metric field in the general case (not for the Einstein universe). This development finds a clear expression in a 1920 paper,\footnote{\citet{Einstein:1920g}, pp.11-13.} where Einstein speaks of the electromagnetic and the gravitational ``ether'' of GR as in principle different from the ether conceptions of Newton, Hertz, and Lorentz. The new, generally relativistic or ``Machian ether'', Einstein says, differs from its predecessors in that it interacts (\emph{bedingt und wird bedingt}) both with matter and with the state of the ether at neighbouring points.\footnote{The fact that Einstein uses the word `ether' extensively in \citet{Einstein:1920g} seems more due to the circumstances in which the paper was delivered than to the content Einstein wanted to bring the term across; Einstein rarely used the term in later writings when referring to $\g$ or $\F$, but sometimes said that whether one says `space-time' or `ether' does not really matter in the end. The fact that Einstein used the term excessively in \citet{Einstein:1920g}, however, must be related to it being his inaugural lecture in Leiden, which Lorentz and Ehrenfest had engineered and which Einstein used to express his reverence for Lorentz.} There can be little doubt that the discovery of the partial dynamical autonomy of the metric field was an unwelcome surprise for Einstein; that as a devotee of Mach he had been reluctant to accept that the metric field was not, in the end, 
``conditioned and determined'' by the mass-energy-momentum $T_{\mu\nu}$ of matter.

But by the time he gave his Princeton lectures in 1921, Einstein had clearly made virtue out of necessity. He had come to see the \textit{inter}active nature of the metric field as a blessing in disguise. This transition in Einstein's thinking was first brought out in 1920, in the course of a correspondence Einstein had with Moritz Schlick.

Schlick started  out as a physicist, and graduated with a PhD on optics in 1904, under the supervision of Max Planck --- maybe the only physicist that Einstein revered as much as Hendrik Lorentz. Schlick then turned to philosophy, and was one of the first philosophers to analyse general relativity; his earliest article stems from 1915.\footnote{See the introduction to \citet{Schlick:GA} for a detailed biography of Schlick, alongside corresponding references.} In 1917, Schlick wrote another paper for the widely read journal \textit{Die Naturwissenschaften}. The article met with so much interest that it was commissioned to be reprinted as a monograph within the same year, with an enlarged and improved second edition being published in 1919. This was of course the year of Eddington's solar eclipse observations and Einstein's rise to international fame, so that another, again enlarged, edition became necessary in 1920. 

In Einstein's first letter to Schlick, he was full of praise of Schlick's 1915 article on relativity theory. He writes:

\begin{quote}

Yesterday I received your paper and have studied it thouroughly already. It is among the best that has been written on relativity to date. From the philosophical perspective, nothing nearly as clear seems to have been written on the subject.\footnote{Einstein to Schlick, 14 December 1915, Vol. 8, Doc. 165, CPAE.} 

\end{quote}

This is the first of at least 26 letters exchanged by Einstein and Schlick between 1917 and the end of 1920 alone. The frequency of correspondence reached its height in 1920, when Schlick was preparing the fourth edition of his book, and when Einstein was struggling with the blow that de Sitter had dealt to Mach's Principle. An important point in the correspondence had to do with a footnote which was present in the first three editions of Schlick's book; it dealt with the way Einstein had criticised the absolute space of Newtonian mechanics in the thought experiment with the two fluid bodies we encountered in the last section. Schlick writes:

\begin{quote}
	One does not have to understand Newtonian theory as taking Galilean space --- after all, an unobservable thing --- for the \emph{cause} of the centrifugal forces. Instead, one might take the talk of absolute space a mere restatement of the bare fact that these forces exist. [...] One does not need to regard absolute rotation as the cause of the ellipsoidal shape [of $S_2$]. Instead one can say: the former is \emph{defined} by the latter.\footnote{\citet{Schlick:1917}, p.178. Translation by DL.}
\end{quote}

Einstein comments on this remark in a letter to Schlick dated 21 March 1917. He writes:

\begin{quote}
Your are right with your criticism on page 178 (footnote) is legitimate. On close examination, the causality requirement is just not sharply defined. The causality requirement can be satisfied to various degrees.  It can only be said that the general th[eory] of r[elativity] has been more successful than classical mechanics in satisfying it. It might be a rewarding task for an epistemologist to think this through carefully.\footnote{Vol. 8, Doc. 314 CPAE.}
	
\end{quote}

Schlick may well have seen this as a challenge posed by Einstein to himself. At any rate, he did think the matter through carefully, and in 1920, while working towards the 4th edition of his book, Schlick published a paper entitled ``Natural-philosophical considerations of the principle of causality''.\footnote{\citet{Schlick:1920kausal}} This paper sparked a new round of discussions between Schlick and Einstein, especially on the question of the differences between NM and GR.\footnote{See \citet{howard1984realism}, \citet{hentschel1986korrespondenz} and \citet{hentschel1990interpretationen}, especially section 4.7.3, for more on the Einstein-Schlick debate.} 

\subsection{Action--reaction to the rescue}

On June 5, 1920, Schlick sent Einstein the paper on the principle of causality, only four days before it was printed in \textit{Die Naturwissenschaften}.\footnote{See Call No. 21-575 EA for the corresponding letter, and Vol. 5, p. 576 CPAE for a summary of the letter.} Einstein answered two days later with detailed comments on the manuscript. One of the issues he focused on is the question of whether the law of inertia has a different causal status in NM as compared to GR:

\begin{quote}

	Furthermore, on the question of the violation of the postulate of causality through the law of inertia. In your little book you rightly emphasised that I went too far in my exposition [of the 1916 thought experiment]. However, I cannot agree with your current position of the state of affairs. I  think it would be correct to say: Newtonian physics has to attribute objective reality to acceleration, independently of the coordinate system. This is only possible if one regards absolute space (i.e., the ether) as something real. Newton does this in a coherent way. It does not matter whether you call that which you have to refer to in order to give reality to acceleration absolute space, ether or preferred coordinate system. What remains unsatisfactory is the circumstance that this something enters \emph{only one way} (\emph{nur einseitig}) into the causal chain. [...] The absolute space of Newton is independent, cannot be influenced, the $\g$-field of general relativity is subject to laws of nature, determined by matter (not only determining).\footnote{Einstein to Schlick, June 7 1920, Vol. 10, Doc. 47 CPAE.} 

\end{quote}

Schlick answers within three days. Interestingly, he does not comment at all on Einstein invoking the action--reaction principle, but reiterates the point he had made in the above mentioned footnote to his book, claiming that it is compatible with everything Einstein had written in his previous letter: 

\begin{quote}
 
\ldots absolute space, which Newtonian mechanics of course has to presuppose, does not have to be considered by the latter as a cause in the sense of the principle of causality. In other words: Newtonian Mechanics does not have to consider inertial resistance in the context of certain kinds of motion as an effect of an absolute acceleration. It can instead take the former as the defining criterion of the latter.\footnote{Schlick to Einstein, 10 June 1920, Vol. 10 Doc. 51 CPAE.}

\end{quote}

So while Einstein takes it as a given that absolute space acts in NM while spacetime \emph{inter}acts in GR, Schlick questions the first point. Unfortunately, just as Schlick did not react to Einstein's new move of invoking the action--reaction principle, Einstein in his answer to Schlick on 30 June, 1920,  does not react to this interesting position either. Instead, he offers an argument of why space has to be considered as absolute in NM:

\begin{quote}

If I consider the equation $$mass \cdot acceleration = force$$ then ``force'' is something absolute (independent of the coordinate system), just like mass, if only the units (including the unit of length) are fixed. Hence, one also has to attribute an absolute status to acceleration. The latter is, with length and time, defined via $\frac{d^2x}{dt^2}$; thus, one must not also define acceleration via the law of inertia. Instead, one will have to choose to also consider $x$ and $t$ as absolute, i.e. physically sensible, quantities. For $t$ this works out, if one does not take into account the difficulty of simultaneity $c$ = practically $\infty$ via a clock; but for $x$ this does not work out. One is forced to attribute a mystical, i.e. empirically inaccessible reality to space.\footnote{Vol. 10, Doc. 67 CPAE.} 

\end{quote}

We leave the reader to ponder over the validity and relevance of this strange argument, which appears to suggest a tension between the Galilean coordinate transformations, or rather the noninvariant transformations for the spatial coordinates, and the absolute nature of the acceleration caused by Newtonian forces. 
More pertinently, Einstein continues by pointing out how things fare better in GR:

\begin{quote}

By the way, physical space possesses reality according to the general theory of relativity, too, but not an independent one; for its properties are completely determined by matter. Space is incorporated into the causal nexus without playing a one-sided role in the causal chain. 


\end{quote}

The second half of the first sentence is also striking, as Einstein had previously recognised that Mach's principle only holds for certain solutions of the Einstein field equations, not for all of them --- but of course, at the time he considered those solutions for which it held as the only physically relevant ones. At any rate, we here see the complete position which would first be presented in the 1921 Princeton lectures: in Newtonian mechanics space acts without being acted upon, while in general relativity it \emph{inter}acts.

In Schlick's answer of 29 August 1920 (i.e. 2 months later), he succumbs to Einstein.\footnote{Vol. 10 Doc. 116 CPAE.} Even though Einstein had not addressed Schlick's alternative position of the status of absolute space in Newtonian mechanics, and Schlick himself failed to explicitly acknowledge the action--reaction principle as the decisive criterion, Schlick announced he was completely convinced by Einstein's argument, and even points out that he would delete the critical footnote from further editions of his book.

\subsection{The observability of the metric field}
\label{S:obsevability}

A noteworthy episode in the Einstein-Schlick correspondence was spurred by a remark in Schlick's 1920 \textit{Die Naturwissenschaften} paper, that from ``an epistemological point of view it is remarkable that the gravitational field does not represent something observable in the same sense as the relative motion of visible bodies''. Einstein replied, in a letter of 7 June of the same year:\footnote{Vol. 10, Doc. 47 CPAE.}

\begin{quote}
It seems to me unjustified to state that gravitational fields should not be regarded as observable in the same sense as masses ... . 
\end{quote}


In turn, Schlick partially concedes the point, but there is a sting in the tail of his reply:

\begin{quote}
I was probably not right with the assertion that a gravitational field was not observable in the same sense as masses. This does apply, at most, in the very rough sense that one may say: I do perceive two objects but not the gravitational field between them. However, it seems to me a debatable point whether in examining Machian thoughts the word `perceptible' may be taken in the roughest sense.\footnote{Schlick to Einstein, 10 June 1920 (Vol. 10, Doc. 51 CAPE).}

\end{quote}

Schlick is correctly wondering what kind of `observability' criterion is operative in Mach's critique of Newtonian absolute space. If it is the ``rough'' one having to do with direct perceivability, then it is unclear how space-time structure in GR is to avoid Machian condemnation. On the other hand, suppose  the relevant notion of observability is a more subtle one: the gravitational field in GR is said to be observable because, like the electromagnetic field in the case of charged bodies, it has an effect on more directly perceivable masses. In this case it is hard to avoid the conclusion that Newtonian space, insofar as it purports to explain inertia, is also observable, and the Machian condemnation of it collapses. It seems that in his correspondence with Schlick, Einstein is adopting something like the second position, abandoning the epistemological flavour of Mach's original critique of absolute space in favour of another tack: that based on the action--reaction principle. If this is correct, it is another example of the plasticity of Einstein's epistemological reasoning in the light of new developments in the physics (in this case prompted by de Sitter's results), and it anticipates Einstein's well-documented turn to a more realist philosophy and a weakening of  his empiricism.\footnote{See  \citet{holton1968mach} and \citet{vanDongen2010einstein}, p.37-40.}

\subsection{Whence AR?} 
\label{S:Whence AR}

As we mentioned in section 3.3 above, Norton argued in 1999 that AR was in the back of Einstein's mind well before 1920, and indeed formed the stimulus of his original Machian tendencies. Here is a further quote from Norton's study:

\begin{quote}
This view of the deficiency of earlier theories [their violating the action--reaction principle] and general relativity's achievement is not one that grew in the wake of Einstein's disenchantment with Mach's principle. Rather, it was present even in his earliest writings beneath the concerns for the relative motion of bodies and the observability of causes.\footnote{\citet{norton1999general}, p. 810}
\end{quote} 

Indeed, this view is consistent with Einstein's 1922 claim, in the quotation found at the end of section 3.4 above, which (questionably) assigns AR as the true source of Mach's own misgivings about the role of  absolute space in Newtonian mechanics. For his part, Norton quotes from one of the 1913 Einstein papers referred to in section 3.3 above, the relevant passage being: 

\begin{quote}
It is a priori to be expected, even though it is not strictly necessary, that inertial resistance is nothing else but resistance against relative acceleration of the body $A$ in question with respect to the totality of all other bodies $B, C$ etc. It is well-known that E. Mach, in his history of mechanics, was the first who defended this position with sharpness and clarity, so that I can only refer to his work. [...] I will call this position the ``hypothesis of the relativity of inertia''. [...] [A] theory in which the relativity of inertia is realised  is more satisfactory than the current theory, for in the latter the inertial system is introduced; its state of motion, on the one hand, is not conditioned by the status of observable objects (and therefore caused by nothing accessible to perception) but, on the other hand, it is supposed to determine the behaviour of material points.\footnote{\citet{Einstein:1913c}, p.1260-1261.}


\end{quote}

But is this an unambiguous enunciation of the AR principle? Note the emphasis, once again, on the epistemological issue of observability and perception, which seems to foreshadow the way Einstein criticises NM in his 1916 review article. Furthermore, it is puzzling that there does not seem to be any other explicit commitment to the AR principle in Einstein's pre-1920 writings, including his effusive letters to Mach.\footnote{It is true that Einstein rejected  his own 1912  scalar field theory (mentioned footnote \ref{F:Newton3} above) when he discovered that it failed to satisfy Newton's third law of motion concerning action--reaction. But this is a case of the existence of \textit{both} action and reaction, which happen not to be equal and opposite, thus giving rise to an unacceptable force-free accelerative phenomenon. As we stressed in section 2, AR is not be to be conflated with Newton's third law, which is a much stronger constraint on the way bodies act on each other. \label{F:Newton3-2}} And as for Einstein's own 1921 interpretation of Mach, this could plausibly be attributed to a mixture of ignorance and opportunism. Finally, and perhaps most importantly, it seems to us that if there was a \textit{single} core belief (of course, there could have been more) underpinning Einstein's adoption of the Machian relativity of inertia, it was probably the equivalence principle, and not AR.\footnote{\citet{Hoefer:1994}, especially in section 1.1, argues that Einstein's conception of the equivalence principle between 1907 and 1912 already contains an ``implicit Machianism''. We hope to clarify in this section how the equivalence principle and the relativity of inertia were likely related to each other in Einstein's mind.} 

In his letters leading up to the paper in which the above quote appears, Einstein keeps emphasising that the fundamental idea on which everything else rests is that gravitational fields and ``acceleration fields''\footnote{Einstein uses this term for example in a letter to Lorentz from 18 February 1912 (Vol. 5 Doc. 360 CPAE) when discussing the alleged equivalence between the two. Note, however, that Einstein puts `acceleration fields' in quotes himself.} are equivalent. Suppose the familiar notion is accepted that gravitation is some kind of interaction between massive bodies, despite the fact that Einstein is embarking on a radically non-Newtonian theory of gravity.  Then if gravitation and acceleration/inertia are to be truly equivalent,  inertia \emph{must} also be an interaction between massive bodies. Indeed, Einstein seems to incorporate such logic in his letter to Mach from 25 June 1913:

\begin{quote}
Next year the eclipse is supposed to show whether light rays are bent by the sun, whether, in other words, the fundamental assumption of the equivalence between acceleration of the frame of reference on the one hand and the gravitational field on the other is indeed right. If it is, then your brilliant investigations on the foundations of mechanics receive --- despite Planck's unjustified criticism --- a shining confirmation. For it follows with necessity that \emph{inertia} has its origin in a kind of \emph{interaction} of the bodies, completely in accordance of your thoughts on Newton's bucket experiment.

\end{quote} 

Note in particular the end of the first sentence and the beginning of the second: \emph{If} the equivalence principle is confirmed, \emph{then} so is the relativity of inertia.

Of course, this inference holds only if \emph{both} the equivalence principle and the claim that gravitation is an interaction hold. There is a short publication in which Einstein replies to criticism from Ernst Reichenb\"acher from  1920,\footnote{\citet{Einstein:1920k}.} the year of his intense exchange with Schlick, where Einstein is explicit in regard to the claim that gravitation is an interaction, with the clarification that the interaction is said to be mediated by $\g$. The outcome, incidentally, is a revised description of the 1916 thought experiment of the two rotating spheres:

\begin{quote}

Mr. \emph{Reichenb\"acher} misunderstood my considerations regarding two celestial bodies rotating with respect to one another. One of these bodies is rotating in the sense of Newtonian Mechanics, and thus flattened by centrifugal effects, the other is not. This is what the inhabitants would measure with rigid rods, tell each other about it, and then ask themselves about the real cause of the different behaviour of the celestial bodies. (This has nothing to do with Lorentz contraction.) \emph{Newton} answered this question by declaring absolute space real, with respect to which one but not the other allegedly rotates. I myself am of the Machian opinion, which in the language of relativity theory can be put in the following way: All masses of the world together determine the $\g$-field, which is, judged from the first celestial body, a different one than judged from the second one; for the motion of the masses producing the $\g$-field differ significantly. Inertia is, in my opinion, a (mediated) interaction between the masses of the world in the same sense as those effects which in Newtonian theory are considered as gravitational effects. 

\end{quote}

A more concise statement of the interaction claim was made in 1921,\footnote{\citet{einstein1921geometrie} p. 12 see also Vol.7, Doc. 31 CPAE for a similar statement from December 1919 / January 1920.}, in the context of providing an argument supporting what we today call the Einstein universe, i.e., the universe in which Mach's Principle would still hold despite de Sitter's result:

\begin{quote}

[I]t seems, therefore, natural to trace back the complete inertia of a body to an interaction between itself and the other bodies in the world, just as since Newton gravitation is traced back completely to an interaction between the bodies.

\end{quote}

To summarise, it seems fair to say that Einstein did not need a variant of the action--reaction principle as a reason to adopt the relativity of inertia in 1913. His strong belief in the equivalence between gravity and inertia, together with his retention of the Newtonian tenet that gravity is an interaction between bodies, could be seen as reason enough.\footnote{Compare \cite{Norton:1989b}, p. 24: ``[I]t was natural for expect that the extended theory, which dealt with general gravitational effects, would explain the observed disposition of inertial frames of reference in terms of the matter distribution of the universe. For the structure that determined this disposition would behave in many aspects like a traditional gravitational field and therefore be strongly influenced by any motion of its sources, the masses of the universe.''} Furthermore, it is the pairing of the equivalence principle and the principle of the relativity of inertia, together with the principle of relativity, that Einstein mentions repeatedly up until 1920 as the cornerstones of GR, whereas AR only really takes centre stage in 1920 in the correspondence with Schlick and in subsequent publications. For these reasons, we are inclined to believe that the 1920 correspondence brought out a watershed in Einstein's thinking, marking an unprecedented shift in Einstein's interpretation of the superiority of GR over preceding theories of space-time: its superiority now rested on satisfaction of the action--reaction principle, rather than implementation of Mach's original analysis of inertia.

\section{Final remarks}

Einstein's frequent references to GR's vindication of the action--reaction principle in the years following his 1921 Princeton lectures  have been noted in a number of studies.\footnote{See in particular \citet{norton1999general} and \citet{Hoefer:1995}.} A particularly telling quotation is from a letter Einstein wrote a year before his death to Georg Jaffe:
\begin{quotation}
You consider the transition to special relativity as the most essential thought of relativity, not the transition to general relativity. I consider the reverse to be correct. I see the most essential thing in the overcoming of the inertial system, a thing which acts upon all processes, but undergoes no reaction. The concept is in principle no better than that of the centre of the universe in Aristotelian physics.\footnote{See \citet{stachel1986einstein}, p. 377.}
\end{quotation} 

For Einstein, the glory of GR rested partly on its alleged superiority to preceding theories of space-time which involve absolute structure. His 1924 essay ``On the ether'' contains a particularly clear denunciation of Newtonian mechanics in terms of its violation of AR.\footnote{\citet{einstein1991uber}, p. 88.} But caution should be exercised when extrapolating backwards, as it were, in the history of physics.  It doesn't automatically follow from the fact that GR satisfies AR, that NM and SR don't, as we mentioned in section 1 above. To repeat, Einstein was content in his 1905 development of SR to explicitly borrow the inertial frames from NM, without any fretting about the correct metaphysics of action. Of course, if AR is to be respected in these theories, inertia must be taken as a brute fact, a position advocated, in different ways, by Schlick and others, as we have seen. Such a position is surely defensible in the context of these theories.

The two epigrammatic Einstein quotations cited at the beginning of this essay underscore how Einstein's thinking changed between 1905 and 1913, and again between 1913 and 1924. In the years 1912 and 1913, when Mach's influence on him may have been greatest, Einstein had convinced himself that the phenomenon of inertia required a causal explanation, while regarding as absurd the notion of immaterial space acting as such a cause. By 1924, he was stressing that the metric field in GR is as real and efficacious as the electromagnetic field, and in particular could indeed be seen as the origin of inertia. (But it is worth stressing here that Einstein did not view GR as furnishing a \textit{geometric} explanation of gravitational phenomena; he continued to reject the notion of space, or space-time, as providing the cause of inertia.\footnote{For details of Einstein's arguments against seeing GR as a `geometrization of gravity', see \citet{Lehmkuhl:geo}; for related arguments, see \citet{anderson1999does} and \citet{brown2009behaviour}.})

Nowadays, acceptance of Einstein's 1924 claim should be seen to rest not simply on the nature of $\g$ and its geodesics, but rather on the so-called \textit{geodesic theorem}, which demonstrates that the form of Einstein's field equations, along, it must be noted, with other plausible universal assumptions about matter fields, imply that the world-lines of test particles are time-like geodesics as defined by the metric field.\footnote{See \citet{malament2012remark}.} Note that the theorem deals with an idealisation; it states that extended, but \textit{truly} freely-falling 
bodies only \textit{approximately} move inertially.\footnote{See \citet{Brown:2007}, p. 141, and particularly \citet{tamir2012proving}.}  In fact, it is a subject worthy of investigation as to whether the details of the theorem are strictly consistent with Einstein's insistence that a violation of AR holds in theories with absolute space-time structure.\footnote{See the brief comments in \citet{Brown:2007}, pp. 141, 142} But such an investigation must be pursued elsewhere. It is our hope that in the present essay, some further light has been shed on the circumstances which led Einstein to bring to the fore the role of the action--reaction principle in his new theory of gravity. 

\section{Acknowledgments}
H.R.B. thanks Partha Ghose, and Peter Ronald deSouza, Director of the IIAS, for the invitation to speak at the 2012 conference ``The Nature of Reality: The Perennial Question". He thanks Julian Barbour for very helpful discussions of the first draft of the paper, in which a number of mistakes were noted and suggestions for improvements were made. He also acknowledges fruitful discussions with Paul Lodge and Gonzalo Rodriguez-Pereyra concerning Leibniz's theory of causation. D.L. thanks the Einstein Papers Project, in particular Diana Kormos Buchwald and Tilman Sauer, for training without which his part in this work would not have been possible, and the Center for Philosophy of Science of the University of Pittsburgh for hospitality during part of the work on the article. He also wishes to thank John Norton in particular, for patient and inspiring discussions on Einstein and the interpretation of Einstein's work. Finally both authors thank John Norton, Tilman Sauer and particularly Oliver Pooley for detailed comments on an earlier draft of the paper which led to a number of improvements.

\nocite{CPAE4}
\nocite{CPAE5}
\nocite{CPAE8}
\nocite{CPAE10}

\bibliography{bibliothek3}

\end{document}